\documentclass[namedreferences]{solarphysics}
\usepackage[optionalrh]{spr-sola-addons} 
\usepackage{graphicx}        
\usepackage{color}           
\usepackage{url}             

\usepackage[pdfborder={0 0 0 },urlcolor=blue]{hyperref}
\ifx \doiurl  \undefined \def \doiurl#1{\href{http://dx.doi.org/#1}{\url{#1}}}\fi
\ifx \adsurl  \undefined \def \adsurl#1{\href{http://adsabs.harvard.edu/abs/#1}{\url{#1}}}\fi

\newcommand{\etal}{{\it et al.}}

\newcommand{\rSun}{{r/R_{\odot}}} 
\newcommand{\RSun}{{R_{\odot}}} 
\newcommand{\eg}{{\it e.g.}}    
\newcommand{\ie}{{\it i.e.}}    

\newcommand{\aap}{    {\it Astron. Astrophys.}}
\newcommand{\aaps}{   {\it Astron. Astrophys. Suppl.}}

\newcommand{\apj}{    {\it Astrophys. J.}}

\newcommand{\mnras}{  {\it Mon. Not. Roy. Astron. Soc.}}
\newcommand{\nat}{    {\it Nature}}

\newcommand{\solphys}{{\it Solar Phys.}}

\begin{document}

\begin{article}

\begin{opening}

\title{The Dynamics of the Solar Radiative Zone}

\author{A.~\surname{Eff-Darwich}$^{1,2}$\sep
        S.G.~\surname{Korzennik}$^{3}$
       }
\runningauthor{A.~Eff-Darwich, S.~Korzennik}
\runningtitle{Rotation Solar Core}

   \institute{$^{1}$ Dept.\ Edafologia y Geologia, Univ.\ La Laguna, 38206, Tenerife, Spain
                     email: \url{adarwich@ull.es} \\ 
              $^{2}$ Instituto de Astrofisica de Canarias, 38205, Tenerife, Spain
                     email: \url{adarwich@iac.es} \\
                                $^{3}$ Harvard-Smithsonian Center for Astrophysics, Cambridge, MA 02138, USA
                     email: \url{skorzennik@cfa.harvard.edu} \\
            }

\begin{abstract}

The dynamics of the solar radiative interior are still poorly constrained by
comparison to the convective zone.  This disparity is even more marked when we
attempt to derive meaningful temporal variations.  Many data sets contain a
small number of modes that are sensitive to the inner layers of the Sun, but
we found that the estimates of their uncertainties are often inaccurate. As a
result, these data sets allow us to obtain, at best, a low resolution estimate
of the solar core rotation rate down to approximately $0.2 \RSun$. {We
present inferences based on mode determination resulting from an alternate
peak-fitting methodology aimed at increasing the amount of observed modes that
are sensitive to the radiative zone}, while special care was taken in the
determination of their uncertainties. This methodology has been applied to MDI
and GONG data, for the whole Solar Cycle 23, and to the newly available HMI
data. The numerical inversions of all these data sets result in the best
inferences to date of the rotation in the radiative region. These results and
the method used to obtain them are discussed. The resulting profiles are shown
and analyzed, and the significance of the detected changes discussed.

\end{abstract}
\end{opening}

\section{Introduction}
     \label{S-Introduction} 

Ground-based helioseismic observations (\eg\ GONG: Harvey \etal, 1996; BiSON:
Broomhall \etal, 2009) and space-based ones (\eg\ MDI: Scherrer \etal,
1995; GOLF: Gabriel \etal, 1995; or HMI: Scherrer \etal, 2012),
have allowed us to derive a good description of the dynamics of the solar
interior (\eg\ Thompson \etal, 2003, Garc\'\i a \etal, 2007, Eff-Darwich
\etal, 2002, 2008, Howe, 2009). Helioseismic inferences have confirmed that the
differential rotation observed at the surface persists throughout the
convection zone. The outer radiative zone ($0.3 < \rSun < 0.7$) appears to
rotate approximately as a solid body at an almost constant rate ($\approx 430$
nHz), {whereas it is not possible to rule out a different rotation rate
for the innermost core ($0.19 < \rSun < 0.3$).}  At the base of the convection
zone, a shear layer --- known as the tachocline --- separates the region of
differential rotation throughout the convection zone from the one with rigid
rotation in the radiative zone. Finally, there is a subsurface shear layer
between the fastest-rotating layer, located at about $0.95 \RSun$, and the
surface.  Of course, this rotation profile is not constant; the time-varying
component of the rotation displays clear variations near the surface (known as
the torsional oscillations), while we see hints of variations at the base of
the convection zone, both being likely related to the driving mechanisms of
the solar-activity cycle.

Our understanding of the dynamics of the solar interior has undoubtedly
improved; however we still need to constrain the rotation profile near the
core and fully analyze the nature of the torsional oscillations.  We still do
not know how thin the tachocline really is and what is keeping it this way.
Understanding the tachocline should help discern if there is a fossil magnetic
field in the radiative zone that prevents the spread of the tachocline (Zahn
\etal, 2007), or an oscillating magnetic field (Forg{\'a}cs-Dajka and
Petrovay, 2001). No purely fluid-dynamics mechanism can explain the tachocline,
resulting in a compelling argument for the presence of a strong magnetic field
(Gough and McIntyre, 1998).

The proper knowledge of the relationship between the solar dynamics and its
structure is not important only in order to understand the present conditions
of the Sun, but also to understand the temporal evolution of our star and
other solar-like stars.  It is usually assumed that the main characteristics
of the dynamics of the Sun were established during its contraction phase
(Turck-Chi{\`e}ze \etal, 2010), hence the Sun was not a rapid rotator
when it entered the Zero Age Main Sequence (ZAMS). The transport of momentum
during the contraction phase might have been carried out by a magnetic field
in the core and the diffusion of this field flattened the rotation profile in
the rest of the radiative zone (Duez \etal, 2010).  In any case,
theories about the mechanisms that drive the solar rotation and its spatial
and temporal variations remain to be tightly constrained by improved
helioseismic inversion results. Better rotation profiles mean not only
improved inversion methodologies but improved estimates of
rotational-frequency splittings.

We present here results derived using an improved inversion methodology that
i) adjusts the inversion grid (over both depth and latitude) based on the
data set and its precision, and ii) solves the inversion problem
iteratively. But first we review recent developments in global mode
characterization (Korzennik, 2008) that allowed us to infer with better
confidence the internal-rotation rate and its time varying patterns. We
describe in detail the inversion methodology and show the resulting profiles.
 
\section{The Data Sets}
\subsection{Introduction}

  We have used rotational frequency splittings determined from fitting data
acquired with five different instruments. Two are ground based: the {\it
Birmingham Solar Oscillations Network} (BiSON) and the {\it Global Oscillation
Network Group} (GONG) and three are on-board spacecraft: the {\it Global
Oscillations at Low Frequencies} (GOLF) and the {\it Michelson Doppler Imager}
(MDI) on board the {\it Solar and Heliospheric Observatory} (SOHO), and the
{\it Helioseismic and Magnetic Imager} (HMI) on board the {\it Solar Dynamics
Observatory} (SDO). For all but the last instrument, the available data sets
span well over a decade of observations.  Table~\ref{tab:datasets} summarizes
what data, from which instrument, and for what time span are included in this
study.

\begin{table}
\begin{tabular}{l|c}
 Instrument           & Time span \\\hline
 BiSON (ground-based) & 01 Jan 1992 -- 31 Dec 2002 \\
 GONG (ground-based)  & 07 May 1995 -- 11 Feb 2011 \\
 GOLF (SOHO)          & 21 May 1996 -- 07 Jun 2007 \\
 MDI (SOHO)           & 01 May 1996 -- 12 Dec 2008 \\
 HMI (SDO)            & 30 Apr 2010 -- 16 Sep 2011 \\
\end{tabular}
\caption{List of instruments and time span from which data sets were used in
  the work presented here.}
\label{tab:datasets}
\end{table}


  The data from these five instruments were fitted with various techniques,
and in some cases the same data were fitted with more than one methodology.
  The GOLF and BiSON data were fitted, using ``Sun-as-a-star'' fitting
techniques, as described by Garcia {\em et al.} (2008) and Broomhall {\em et
al.} (2009), respectively. 
The fitting, in both cases, is limited by these instruments' lack of spatial
resolution to low degree modes ($\ell \le 3$).

  The methodology for the mode fitting pipeline used by the GONG project is
described by Anderson {\em et al.} (1990). It processes 108-day long
overlapping time series, each 36 days apart, and individually fits each
multiplet. It does it without including any {spatial} leakage matrix
information and uses a symmetric profile for the mode power-spectral
density. When resolved, spatial leaks are independently fitted, but when they
are not resolved (in most cases), blended leaks are fitted as a single
peak. Since there is no inclusion of any leakage information, the blending
affects the result, skewing the mode frequency and the mode line width.

  The mode-fitting pipelines used by the MDI team (both the standard and
``improved'' pipelines) fit non-overlapping 72-day long epochs. That fitting
methodology fits singlets, using a polynomial expansion in $m$ to model the
frequency splitting, and includes the leakage matrix information (as described
by Schou, 1992). The improved pipeline (Larson and Schou, 2008) includes an
improved spatial decomposition, where the effective instrument plate scale and
our best model of the image distortion is included, as well as an improved
leakage computation that incorporates the distortion of the eigenmodes by
differential rotation (Woodard, 1989). The improved pipeline is set up to fit
either a symmetric or an asymmetric mode power spectral density profile.

\subsection{Our Alternate Peak-Fitting Method}

Korzennik (2005, 2008) has developed and implemented an alternative fitting
methodology, which has processed GONG, MDI, and HMI data. The key elements of
this method are as follow: it fits individual multiplets, simultaneously for
all the azimuthal orders while including the leakage information. The leakage
matrix includes the effect of the distortion of the eigenmodes by differential
rotation (Woodard, 1989). The spectral estimator is a sine multi-tapered one,
whose number of tapers is adjusted to be {\em optimal}, a value derived from
the mode line-width. The mode power-spectral density profile is asymmetric,
the procedure is iterative so as to include mode contamination (mode with a
different radial order [$n$] present in the fitting window), and it includes a
rejection factor, where modes with too low an amplitude are not fitted.

The other major difference in the implementation of this method is that we
choose to fit time series of varying lengths. The gain in signal-to-noise
ratio when using longer time series allows us to derive more accurate mode
parameters, while trading off precision for temporal resolution. We used
$64\times, 32\times, 16\times, 8\times, 4\times$ and $2\times72$-day long,
overlapping, time-series, as well as $1\times72$-day long non-overlapping
epochs {(note that the longer segments all start on 01 May 1996, \ie\ 
the start of science quality observations for MDI)}. This extensive analysis
of some 13 years of data was carried out on the Smithsonian Institution High
Performance Cluster.

This method was used to fit GONG time series, using a leakage matrix
specifically computed for that instrument, although the change in leakage
resulting from the 2001 camera upgrade was not yet included (Schou, private
communication, 2003).  That same method was used to fit MDI data, for the
exact same epochs, but using an MDI-specific leakage matrix. In fact, we
fitted the data using a leakage matrix supplied by the MDI team, as well as
our own independent leakage matrix computation. We used the ``improved'' MDI
time series, where the spatial decomposition includes the effective instrument
plate scale and our best model of the image distortion. We also fitted HMI
{\em provisional} time series (as the HMI processing pipeline is yet to be
finalized). The HMI instrumental image distortion and precise plate scale are
included at the filtergram processing level, and the data were fitted using a
{\em provisional} leakage matrix (\ie\ the one derived for the full-disk MDI
observations).

\begin{figure}[!t]
\begin{center}
\includegraphics[width=.95\columnwidth]{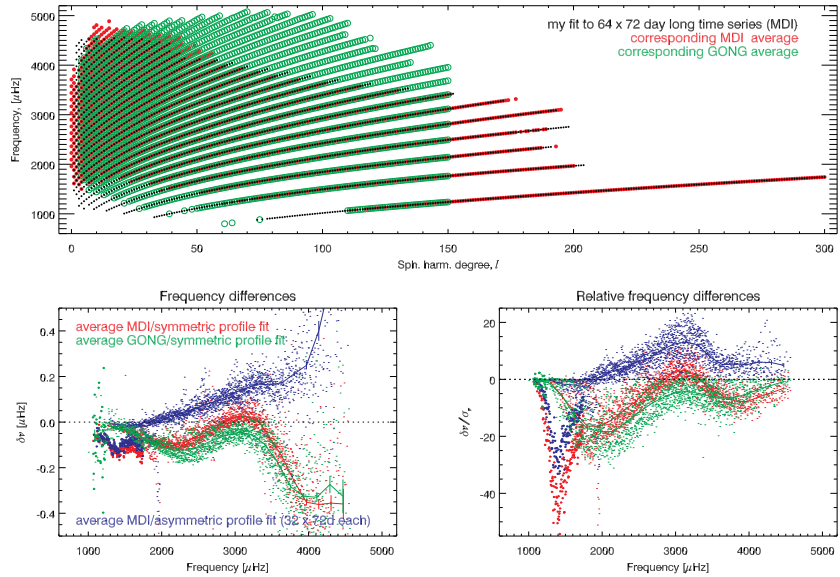}
\end{center}
\caption{Comparison of fitted frequencies (singlets): top panel shows the
coverage in the $\ell-\nu$ plane, bottom panels shows frequency and relative
frequency differences ({\em i.e.}, differences divided by the uncertainty),
{relative to the alternate fitting methodology}. Black dots correspond to
modes fitted using the alternate fitting methodology, applied to
$64\times72$-day long time series, the red and green dots correspond to MDI
and GONG pipeline fitting respectively, while the blue dots correspond to MDI
``improved'' fitting, using an asymmetric profile.  The large dots correspond
to the {\it f}-modes, the curves are the {\it p}-mode corresponding binned
values.}
\label{fig:compare-freqs}
\end{figure}

\begin{figure}[!t]
\begin{center}
\begin{picture}(0,0)
\put( 10,210){(a)}
\put(170,210){(b)}
\put( 10,90){(c)}
\put(170,90){(d)}
\end{picture}
\includegraphics[width=.95\columnwidth]{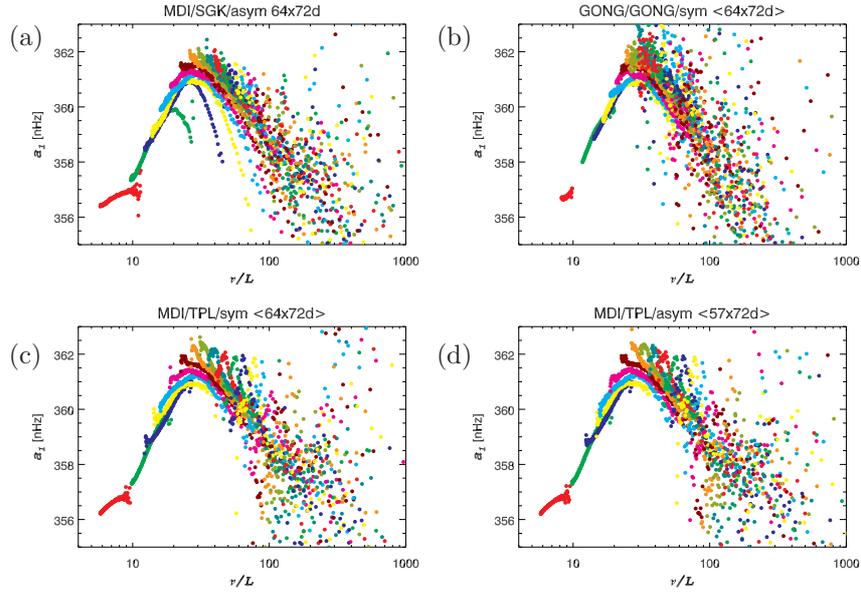}
\caption{Comparison of the frequency splitting leading Clebsch--Gordan
coefficient derived from four mode-fitting procedures: (a) results from using
our alternative fitting methodology, (b) GONG pipeline, (c) MDI improved
symmetric fit, and (d) MDI improved asymmetric fit. {The resultiong $a_1$
coefficients are plotted versus $\frac{\nu}{L}$, where $L^2 = \ell(\ell+1)$,
while the symbol's color corresponds to the mode order, $n$.}}
\label{fig:compare-cgs-1}
\end{center}
\end{figure}

Figures~\ref{fig:compare-freqs} and \ref{fig:compare-cgs-1} and
Table~\ref{tab:compare-cgs} compare results from fitting GONG and MDI data
with the respective projects analysis pipelines and the above described
alternate fitting method. {The table lists the mean and standard deviation
of the differences in the $a_1$ Clebsch--Gordan coefficients (linear term)
estimated by various fitting procedures.}  The resulting singlets show
systematic differences, that are not simply explained by the inclusion or not
of an asymmetric profile, with even larger and systematic differences for the
{\it f}-mode. The comparisons of the rotational-splitting coefficients show
less of a scatter for the linear term, when using the alternate peak-fitting
method, and differences at the few $\sigma$ level.

\begin{figure}[!t]
\begin{center}
\includegraphics[width=.95\columnwidth]{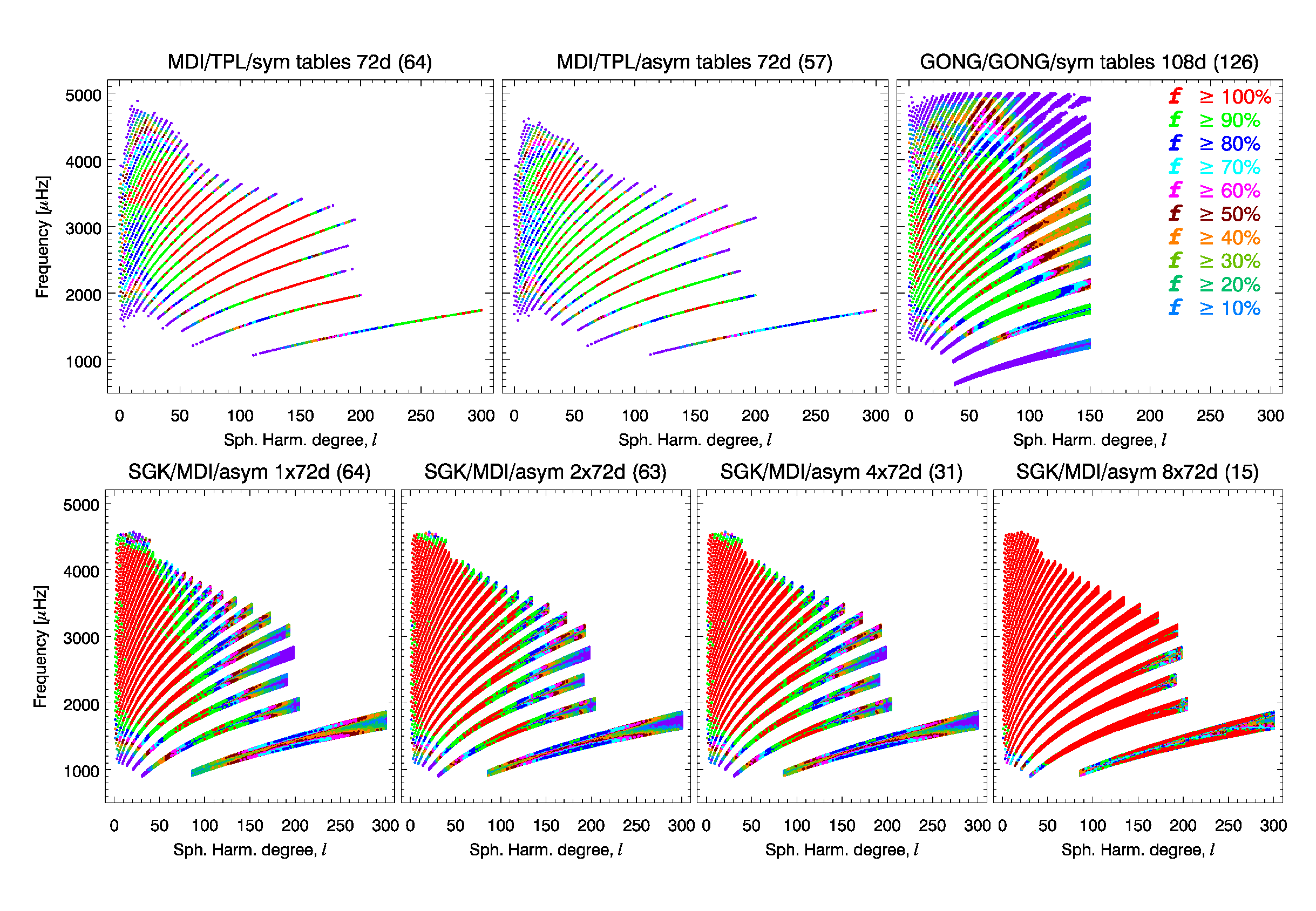}
\caption{Mode attrition in the $\ell-\nu$ plane. The color represents how
often a mode is fitted, with red indicating all the time ($f=100\%$), green 90\%,
etc. The top panels correspond to MDI improved symmetric and asymmetric fit
and GONG pipeline fit. The bottom panels correspond to our alternative fitting
methodology, for $1\times$, $2\times$, $4\times$, $8\times72$-day long time
series (left to right).}
\label{fig:show-attrition}
\end{center}
\end{figure}

But also, if not more important, is the difference in mode attrition, when
using the various fitting methods. Figure~\ref{fig:show-attrition} illustrates
that mode attrition, \ie\ how often a mode is successfully fitted for each
epoch analyzed. That figure shows clearly that the project pipeline methods
produce large attrition, while the alternate peak-fitting method results
display a more consistent fitting pattern. {In order to be confident that
we deduce significant changes of the solar rotation, when inverting rotational
frequency splittings for various epochs, we ought not to inject changes
resulting from using different mode sets in the inversions.} The estimated
solutions of an inversion problem are some weighted spatial average of the
``real'' underlying solution. Those weights (also known as resolution kernels)
depend on the extent of the input set, and thus change when the input sets
change.

\begin{table}[!t]
\begin{tabular}{rrr}
& \multicolumn{1}{c}{$\delta a_1$}        
& \multicolumn{1}{c}{$\delta a_1/\sigma_{a_1}$} \\
& \multicolumn{1}{c}{[nHz]} & \\\hline
GONG (sym.)  {\it vs} alternate (asym.) $64\times72$-day long & $-0.277 \pm 0.984$ & $-0.917 \pm  1.279$ \\
MDI  (sym,)  {\it vs} alternate (asym.) $64\times72$-day long & $ 0.051 \pm 0.635$ & $ 0.534 \pm  2.888$ \\
MDI  (asym.) {\it vs} alternate (asym.) $32\times72$-day long & $ 0.096 \pm 0.769$ & $ 1.398 \pm  2.384$ \\
\end{tabular}
\caption{Comparison of resulting $a_1$ Clebsch--Gordan coefficients (linear
term) derived from four mode fitting procedures. {The table lists the mean and standard deviation
of the differences in $a_1$}.} \label{tab:compare-cgs}
\end{table}

\section{Inversion Methodology}

  The starting point of all helioseismic, linear rotational inversion
methodologies is the functional form of the perturbation in frequency [$\Delta
\nu _{n \ell m}$] induced by the rotation of the Sun, $\Omega(r,\theta)$:
\begin{equation} 
\Delta \nu _{n \ell m} = \int_0^{\RSun}\!\!\!\! \int_0^{\pi}\!\!
K_{n \ell m}(r,\theta)\,\Omega(r,\theta)\,\mathrm{d}r\, \mathrm{d}\theta \pm \epsilon_{n \ell m}
\label{eq:equation1} 
\end{equation}

  The perturbation in frequency [$\Delta \nu _{n \ell m}$] with the
observational error [$\epsilon_{n \ell m}$] that corresponds to the rotational
component of the frequency splittings, is given by the integral of the product
of a sensitivity function, or kernel  [$K_{n \ell m}(r,\theta)$] with the
rotation rate [$\Omega(r,\theta)$] over the radius [$r$] and the co-latitude
[$\theta$]. The kernels [$K_{n \ell m}(r,\theta)$] are known functions of the
solar model.

  Equation (\ref{eq:equation1}) defines a classical inverse problem for the
Sun's rotation. The inversion of this set of $M$ integral equations -- one for
each measured $\Delta \nu _{n \ell m}$ -- allows us to infer the rotation rate
profile as a function of radius and latitude from a set of observed rotational
frequency splittings (hereafter referred as splittings).

Our inversion method requires the discretization of the integral
relation to be inverted. In our case, Equation~(\ref{eq:equation1}) is transformed
into a matrix relation
\begin{equation}
  D = A x + \epsilon \label{eq:equation2}
\end{equation}
where $D$ is the data vector, with elements $\Delta \nu _{n \ell m}$ and
dimension $M$, $x$ is the solution vector to be determined at $N$ model grid
points, $A$ is the matrix with the kernels of dimension $M \times N$, and
$\epsilon$ is the vector containing the corresponding observational
uncertainties. The number and location of the $N$ model grid nodes are
calculated according to the effective spatial resolution of the inverted data
set. Such a procedure produces a non-equally spaced (\ie\  unstructured) mesh
distribution. A complete description and examples of the gridding methodology
can be found in Eff-Darwich and P\'erez-Hern\'andez (1997) and Eff-Darwich
{\em et al.} (2010).

The resulting unstructured grid is used to compute the matrix $A$ in
Equation~(\ref{eq:equation2}). That equation is then solved with a modified version
of the iterative method developed by Starostenko and Zavorotko (1996). This
approach calculates $x$ according to the following algorithm:
\begin{equation}
x^{k+1} = x^k - B^{-1}A^TR^{-1}(Ax^k-D)  
\end{equation}
where $k$ is the iteration index. The diagonal matrices $B$ and $R$ are
calculated from the summation of columns and rows of matrix $A$,
respectively. For each iteration, values for the error propagation and data
misfit, $\chi^2 = |Ax-D|^2$, are calculated.

\section{Results}

\begin{figure}[!ht]
\begin{center}
\includegraphics[width=.49\columnwidth]{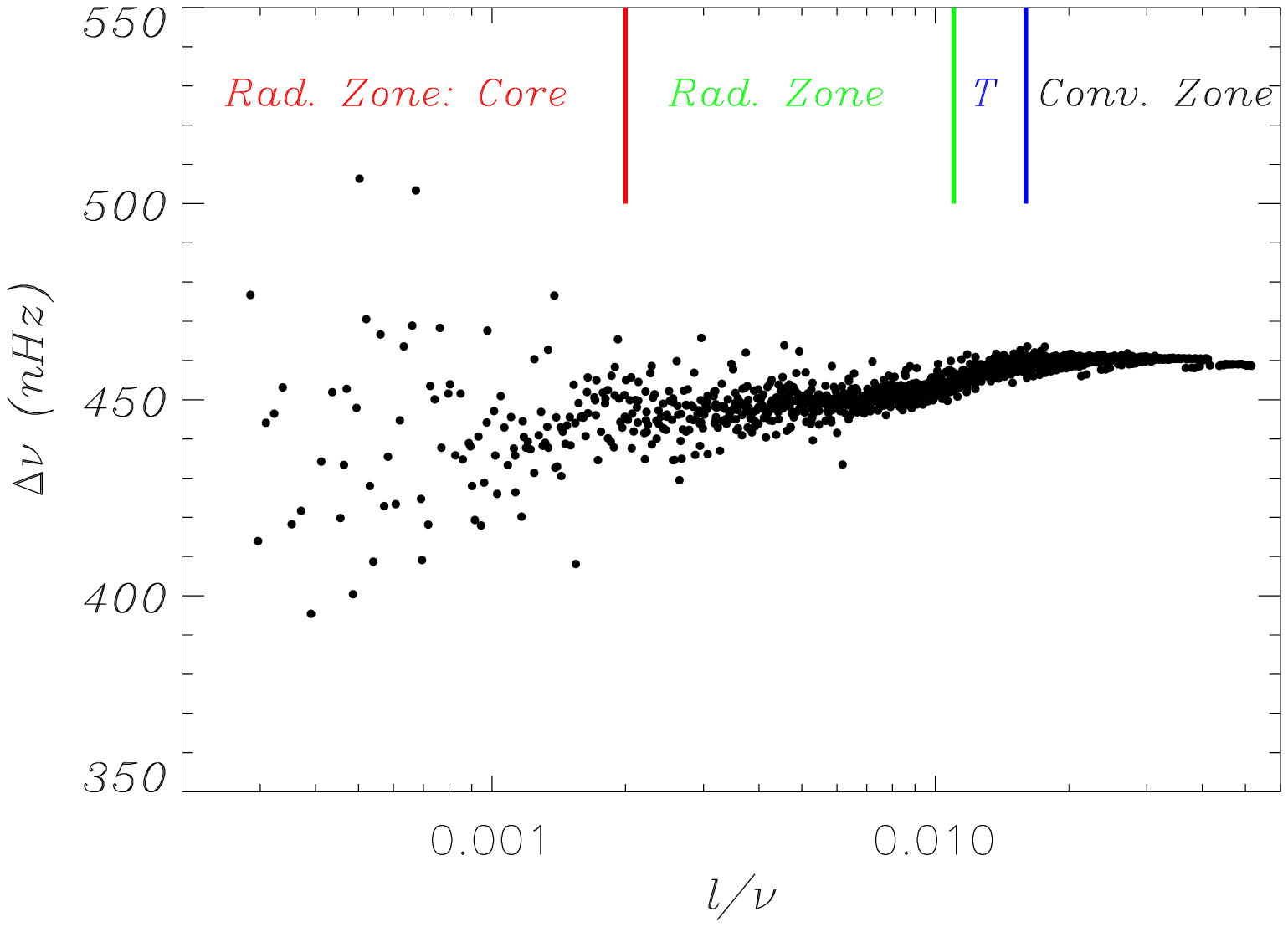}
\includegraphics[width=.49\columnwidth]{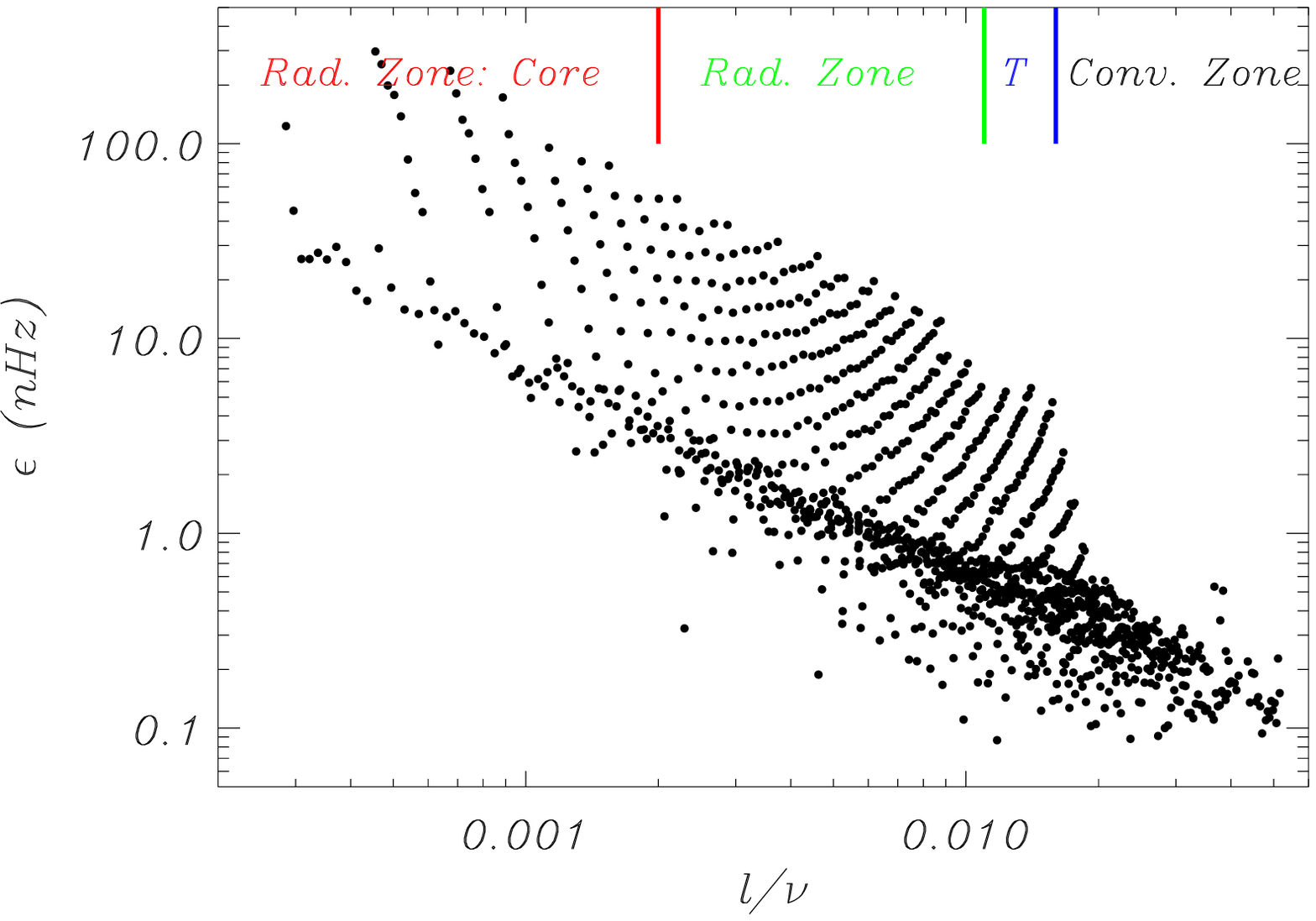}
\caption{Left panel: observational sectoral frequency splittings (MDI
$64\times 72$-day long time series) as a function of {the $\frac{\ell}{\nu}$
ratio, a proxy for} the inner turning radius. For illustrative purposes, the
approximate extent of the solar core, radiative zone, tachocline and
convective zone are represented. Right panel: as in the left panel, but for
the observational uncertainties of sectoral frequency splittings. }
\label{fig:0}
\end{center}
\end{figure}

Helioseismology, as a tool to infer the properties of the solar interior, is
based on the fact that different mode sets are sensitive to different layers
of the Sun. Hence, by combining these mode sets, it is possible to derive the
structure and dynamics of the solar interior. However, these sets are not
homogeneous and the number and quality of the modes that are sensitive to the
solar radiative interior is significantly lower than those sensitive to the
convective zone and the surface layers (as shown in Figure~\ref{fig:0}).
Therefore, the dispersion and the level of uncertainties of the modes that are
sensitive to the core are the largest for the entire data set. Another problem
arises as we look closely at the uncertainties (see Figure~\ref{fig:0}): the
error level as a function of radius is not strictly monotonic. For a given
inner turning radius, the scatter of the errors is rather large, and is
primarily the consequence of the reduced accuracy of estimates at high
frequencies.

\begin{figure}[!ht]
\begin{center}
\includegraphics[width=.49\columnwidth]{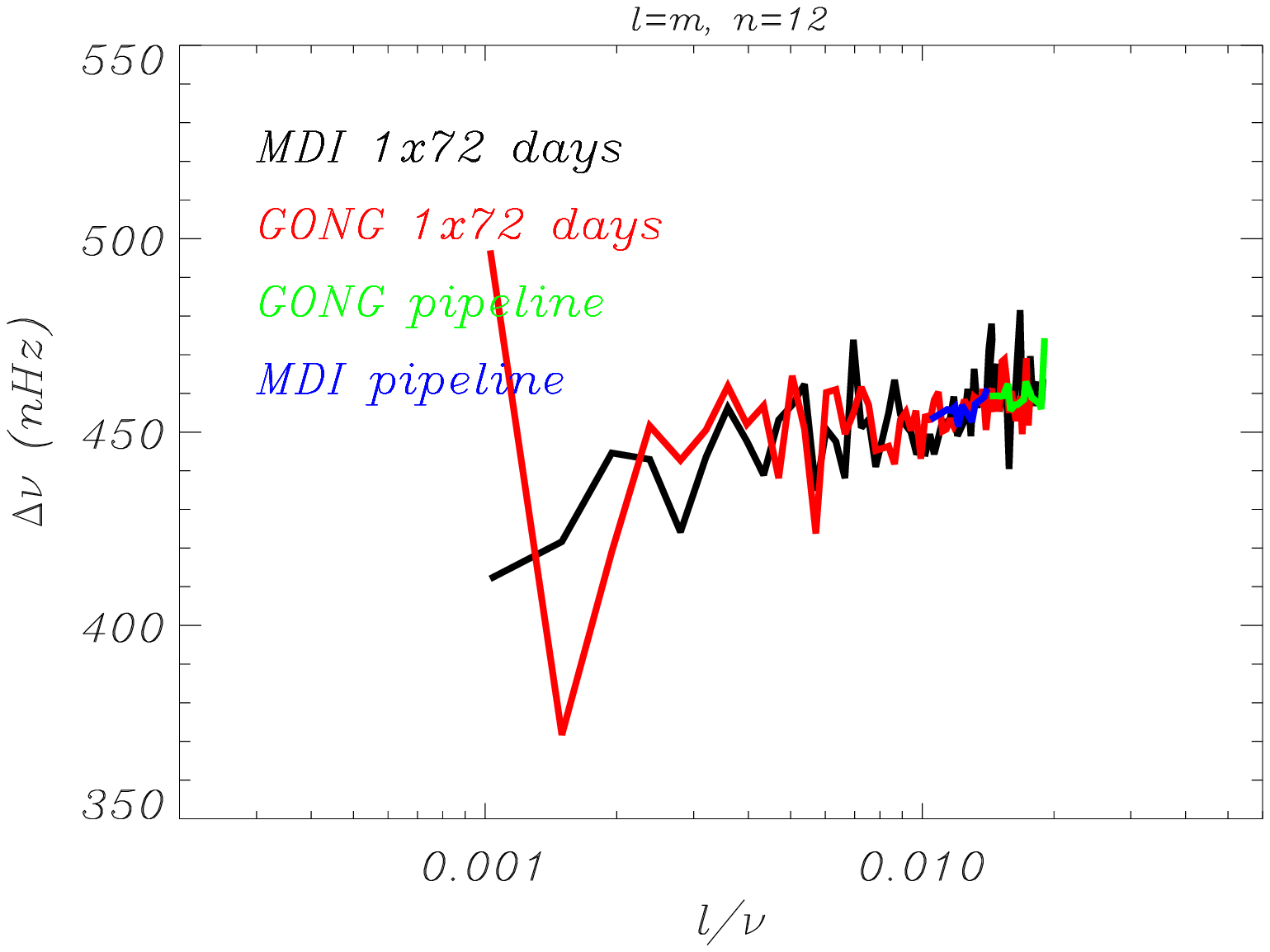}
\includegraphics[width=.49\columnwidth]{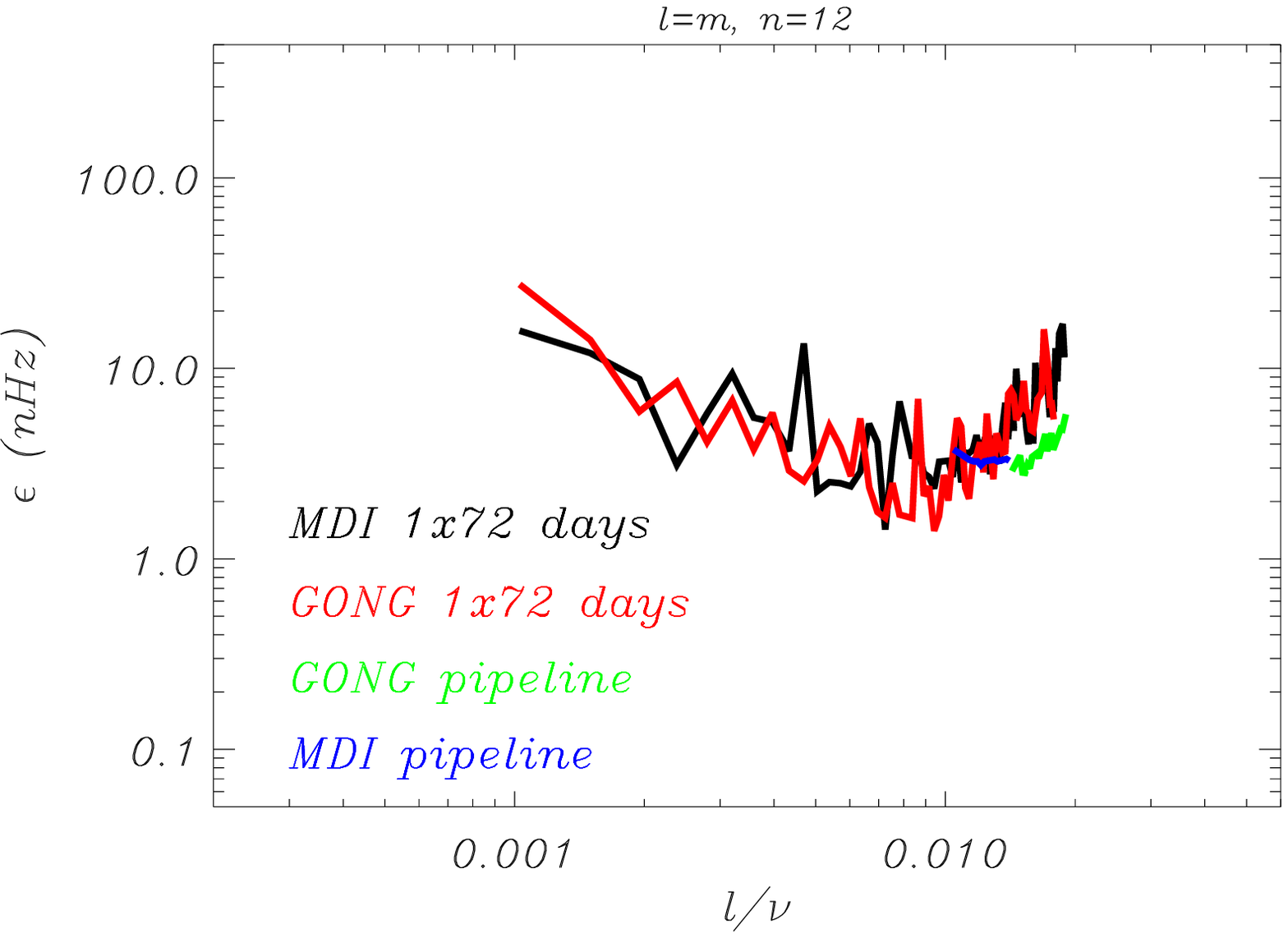}
\caption{Left panel: observational sectoral frequency splittings as a function
of {the $\frac{\ell}{\nu}$ ratio, a proxy for the inner turning radius},
for $n=12$ modes obtained by fitting MDI $1\times 72$, GONG $1\times 72$-day long time-series
and the MDI and GONG team pipelines.  Right panel: as in the left panel, but
for the observational uncertainties. }
\label{fig:17}
\end{center}
\end{figure}

\begin{figure}[!ht]
\begin{center}
\includegraphics[width=.49\columnwidth]{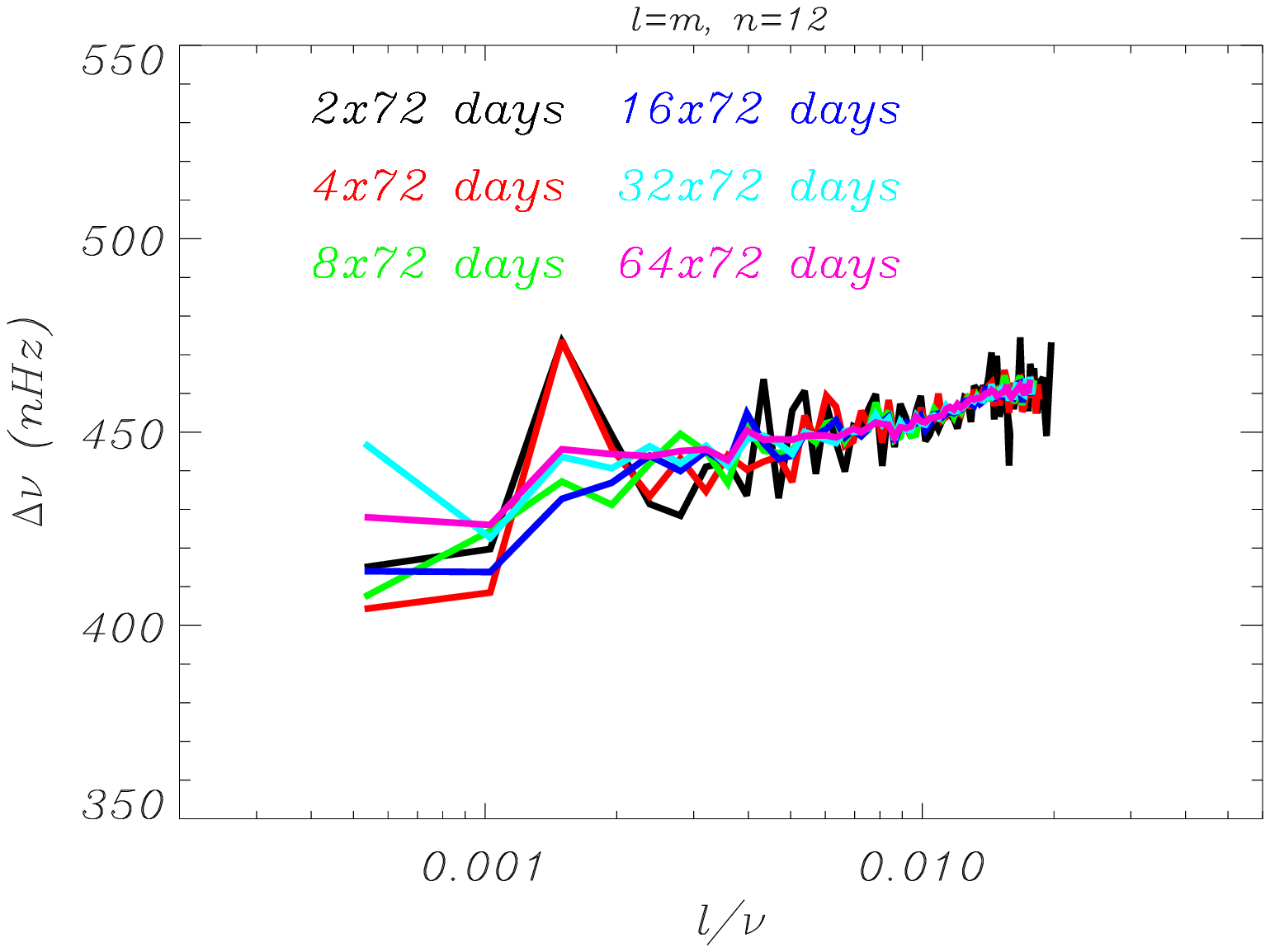}
\includegraphics[width=.49\columnwidth]{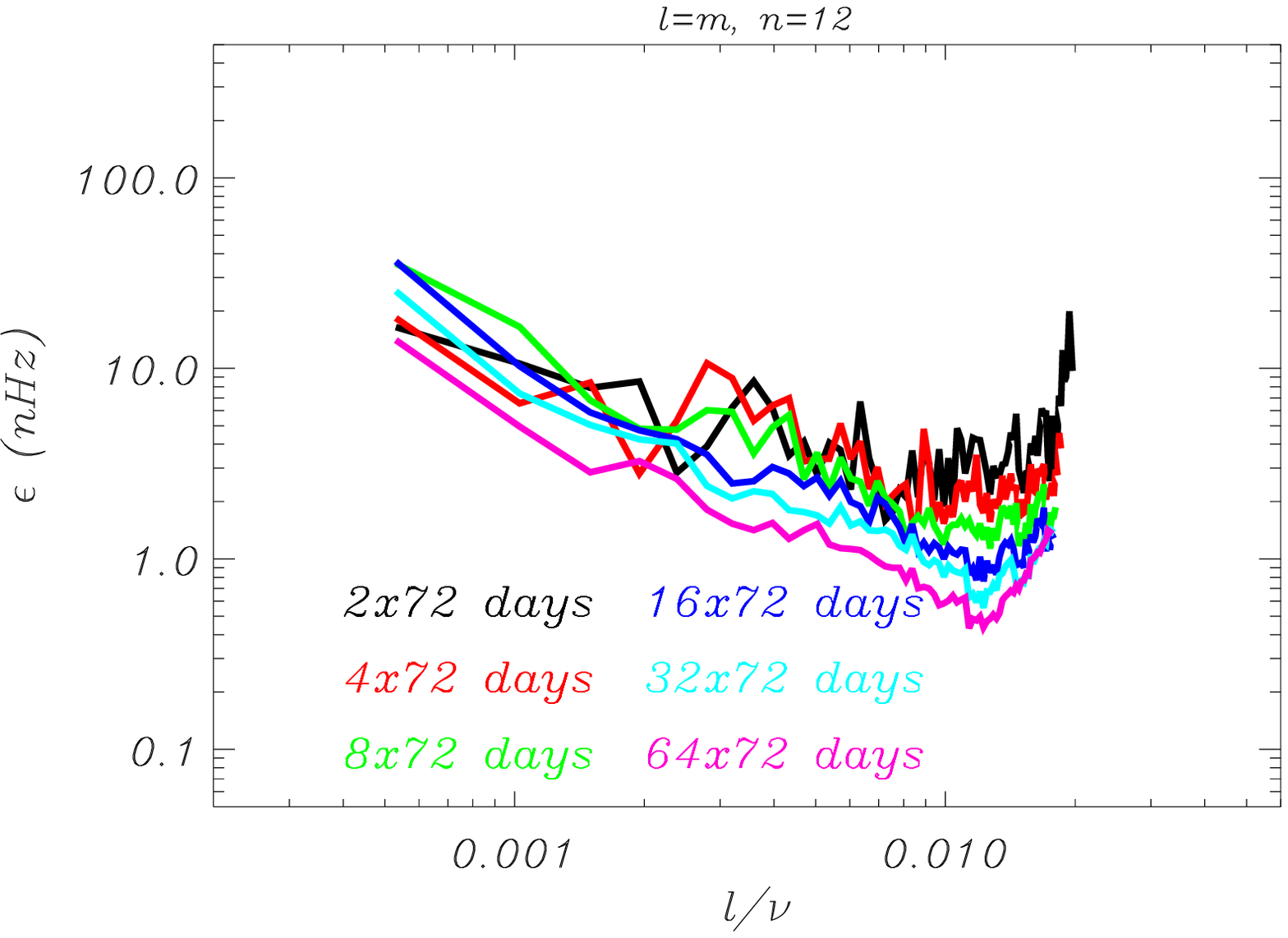}
\caption{Left panel: observational sectoral frequency splittings as a function
of {the $\frac{\ell}{\nu}$ ratio, a proxy for the inner turning radius},
for $n=12$ modes obtained by fitting MDI
$2\times$, $4\times$, $8\times$, $16\times$, $32\times$, and $64\times 72$-day
long time-series. Right panel: as in the left panel, but for the
observational uncertainties. }
\label{fig:15}
\end{center}
\end{figure}

Figure~\ref{fig:17} shows how consistent both the range in degree and
frequency are when the alternative fitting technique developed for this work
is used on data from different instruments.  By contrast, the mode sets
obtained by the team pipelines, for both GONG and MDI, differ significantly,
especially for their frequency spans.  The consistency of our fitting
technique is shown in Figure~\ref{fig:15}: this figure shows how both the
uncertainties and the data dispersion are reduced when the length of the
time series analyzed is increased.  The improvement is less apparent for the
modes that are more sensitive to the solar core and in the data sets
corresponding to shorter time-series. However, in the case of the $64\times
72$-day long mode set, the uncertainties for the modes sensitive to the core
are reduced by a factor of four relative to the mode sets obtained from shorter
time series.

\begin{figure}[!ht]
\begin{center}
\includegraphics[width=.8218\columnwidth]{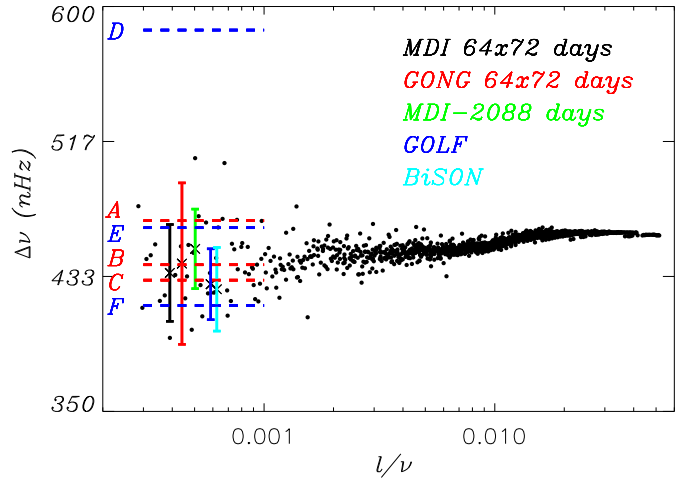}
\caption{Observational sectoral frequency splittings as a function of the
horizontal phase speed for $\ell=1$ modes (MDI $64\times 72$-day long time
series). The vertical colored lines represent the frequency-averaged (1.1to 3.3
mHz) $\ell=1$ rotational splittings derived from our peak-fitting methodology
(MDI and GONG $64\times 72$ days), the MDI pipeline (using a 2088-day long
time series), GOLF and BiSON.The red dashed lines represent the average values
of the theoretical $\ell=1$ rotational splittings, if the radiative zone were
rotating rigidly at a rate of $432$ nHz, with rates below $0.12R_\odot$ of
$2832$, $835$, and $132$ nHz (A, B and C respectively). The blue dashed lines
represent the average values of the theoretical $\ell=1$ rotational
splittings, with rates below $0.2R_\odot$ of $2832$, $835$, and $132$ nHz (D, E
and F respectively).  }
\label{fig:3}
\end{center}
\end{figure}

Hence, uncertainties of the data sensitive to the solar core rotation decrease
when longer time-series and better fitting technique are used.  The level of
uncertainties that we need to reach to counteract the low sensitivity of the
modes to these regions is illustrated in Figure~\ref{fig:3}, with test
profiles. Two sets are presented: i) one set where the radiative zone is
rotating rigidly, at a rate of $432$ nHz, and below $0.12R_\odot$ at rates of
$2832$, $835$, and $132$ nHz; ii) the other set where the radiative zone is
also rotating rigidly at a rate of $432$ nHz, and where below $0.2R_\odot$ the
rates are again set to $2832$, $835$, and $132$ nHz.  Out of these six test
profiles, only one is substantially and significantly different from the
frequency-averaged $\ell=1$ rotational splittings (\ie\  averaged over
frequencies in the 1.1 to 3.3 mHz range) derived from our peak-fitting
methodology (MDI and GONG $64\times 72$ days), the MDI pipeline (using a
2088-day long time series), GOLF, and BiSON data sets.

\begin{figure}[!ht]
\begin{center}
\includegraphics[width=.8218\columnwidth]{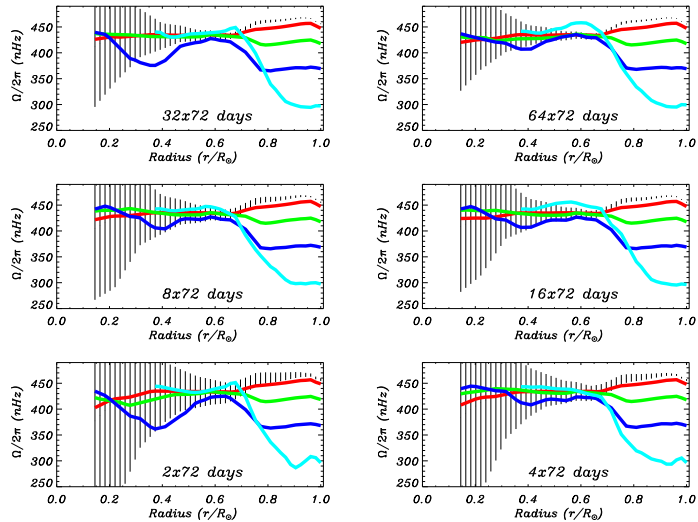}
\caption{Time-averaged rotational profiles obtained from the inversions of
rotational frequency-slittings resulting from fitting MDI $2\times$,
$4\times$, $8\times$, $16\times$, $32\times$, and $64\times 72$-day long
time-series.  Black, red, green, dark-blue, and light-blue lines correspond to
the rotational rate at different latitudes, namely $0, 20, 40, 60$, and
$80^{\rm o}$, respectively. Vertical lines represent the error bars for the
rotational rate at the Equator.}
\label{fig:27}
\end{center}
\end{figure}

\begin{figure}[!ht]
\begin{center}
\includegraphics[width=.8218\columnwidth]{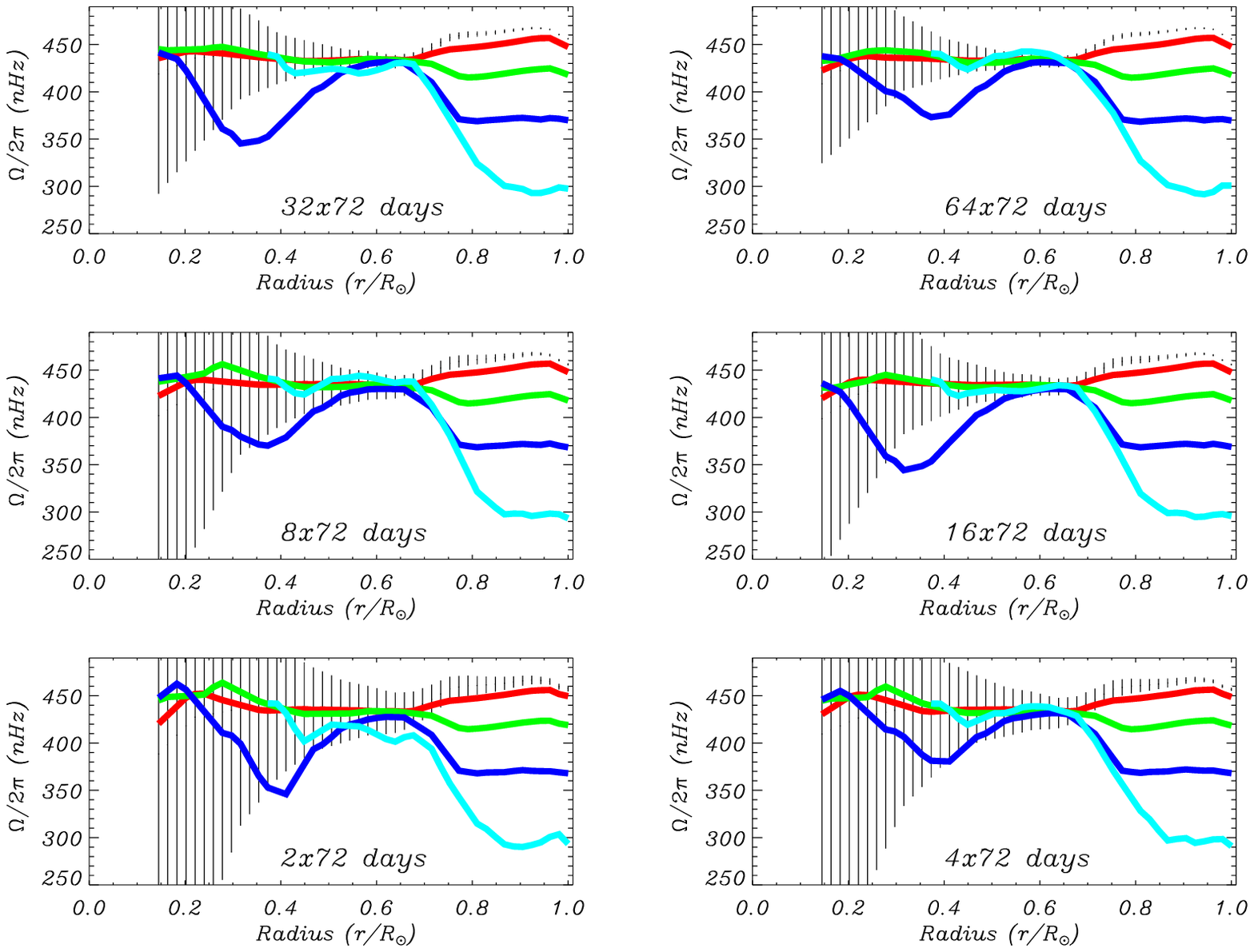}
\caption{Time-averaged rotational profiles obtained from the inversions of
rotational frequency-slittings resulting from fitting GONG $2\times$,
$4\times$, $8\times$, $16\times$, $32\times$, and $64\times 72$-day long
time-series.  Black, red, green, dark-blue, and light-blue lines correspond to
the rotational rate at different latitudes, namely $0, 20, 40, 60$, and
$80^{\rm o}$, respectively. Vertical lines represent the error bars for the
rotational rate at the Equator}
\label{fig:28}
\end{center}
\end{figure}

\begin{figure}[!ht]
\begin{center}
\includegraphics[width=.8218\columnwidth]{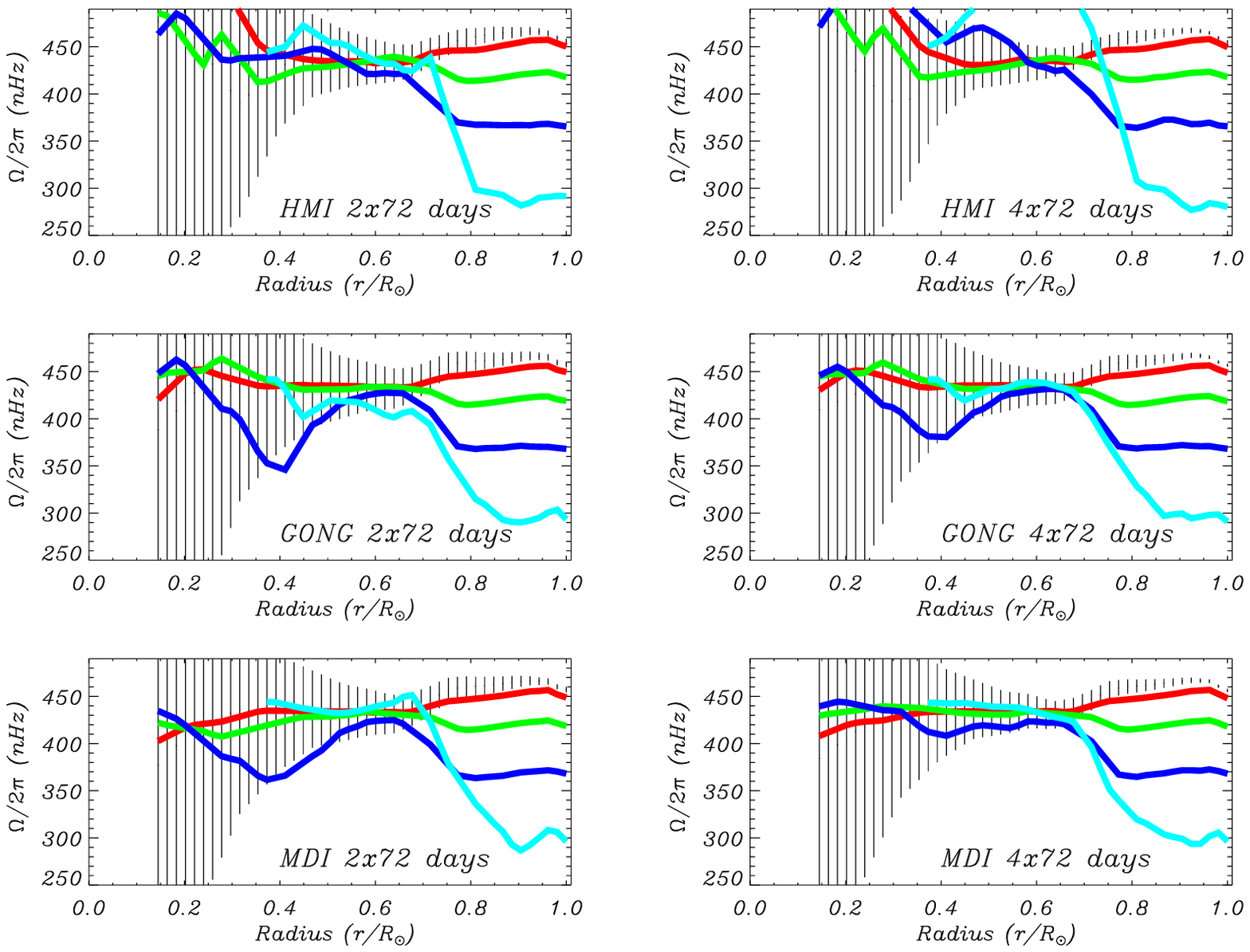}
\caption{Time-averaged rotational profiles obtained from the inversions of
MDI, GONG, and HMI $2\times 72$ and $4\times 72$-day long sets.  Black, red,
green, dark-blue, and light-blue lines correspond to the rotational rate at
different latitudes, namely $0, 20, 40, 60$, and $80^{\rm o}$,
respectively. Vertical lines represent the error bars for the rotational rate
at the Equator.}
\label{fig:30}
\end{center}
\end{figure}

The diagnostic potential of the new global-mode fitting technique when
combined with the improved inversion methodology is illustrated in Figures
\ref{fig:27}, \ref{fig:28}, and \ref{fig:30}, where we present the
time-averaged rotation profiles of the Sun from the surface down to
$0.15\RSun$ that were calculated, using either MDI or GONG, and $2\times$,
$4\times$, $8\times$, $16\times$, $32\times$, and $64\times 72$-day long
time-series.  Inversions using recent HMI data ($2\times$ and $4\times 72$-day
long) are also presented, although they are not yet comparable to either MDI
or GONG results, since the amount of HMI observations is still significantly
smaller.

Both MDI and GONG inversions give similar results, with the largest
discrepancies at high latitudes and below $0.40R_\odot$. {The most
significant difference between the inversions obtained by the same instrument
is the reduction of the uncertainties, in particular random noise, when the
length of the time-series used to fit is increased. This reduction is
particularly important in the inner radiative core.}


All results are compatible with a radiative zone rotating rigidly at a rate of
approximately 431 nHz; however, it is not possible to disregard a faster or
slower rotator below $0.2\RSun$ (\ie\  up to 600 or down to 300 nHz). Although
the radiative zone seems to rotate rigidly, there is a consistent and
systematic dip in the rotation profile located at approximately $0.4\RSun$ and
$60^{\rm o}$ in latitude. This dip is seen in both MDI and GONG results,
notwithstanding the actual length of the fitted time series.

\begin{figure}[!ht]
\begin{center}
\includegraphics[width=.48\columnwidth]{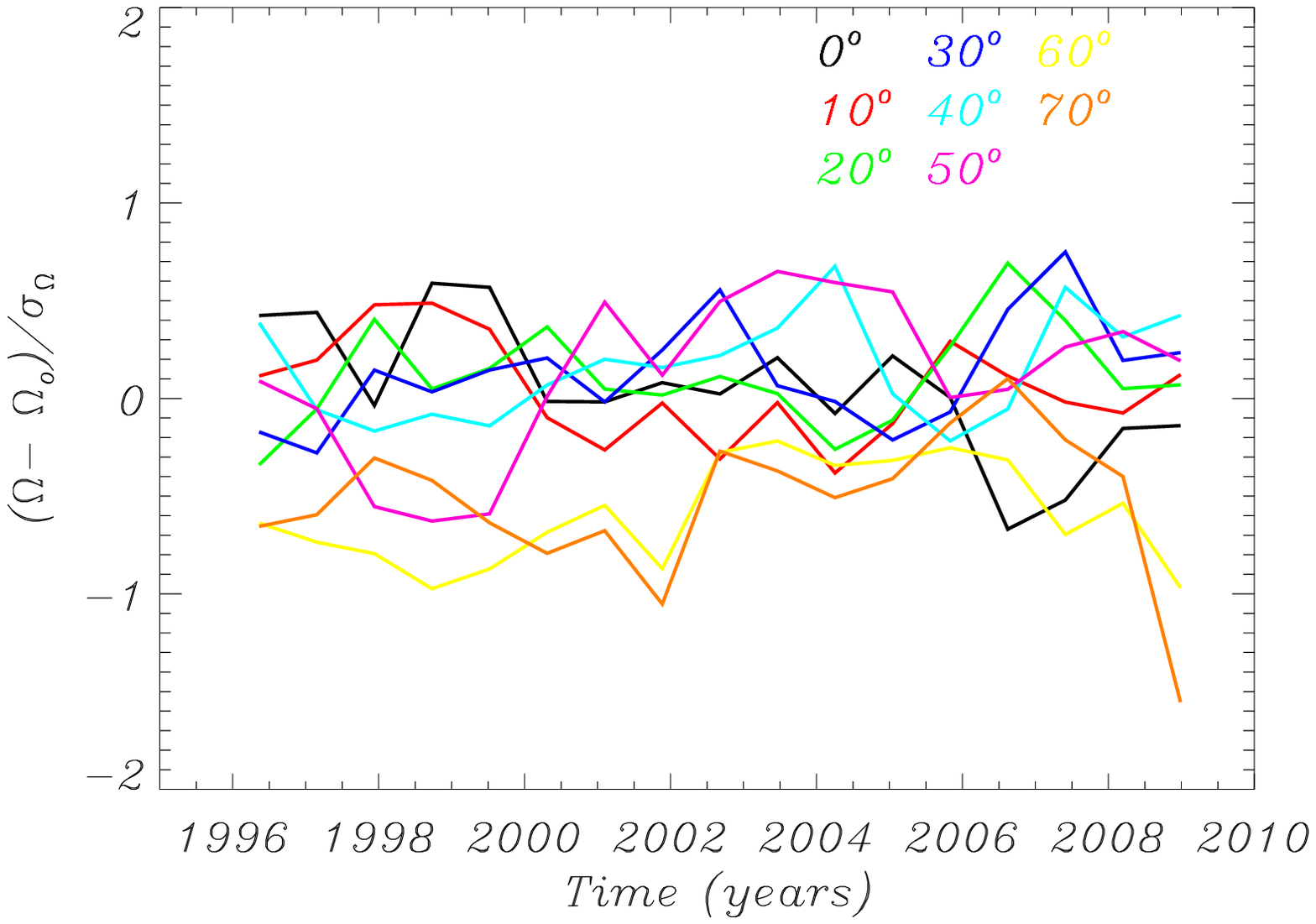}
\includegraphics[width=.48\columnwidth]{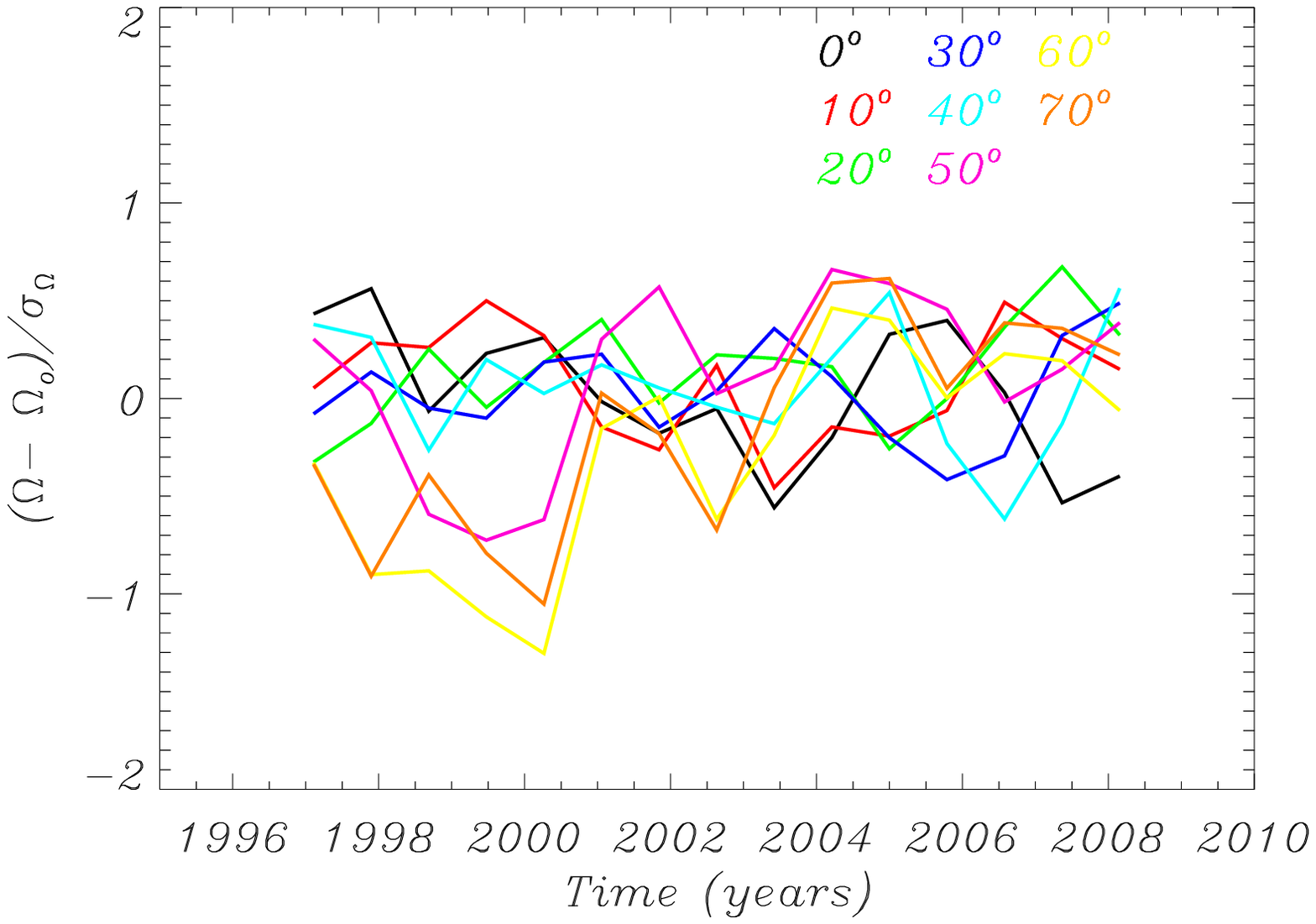}
\caption{Left panel: temporal evolution of the {relative residual rotation
rate, $(\Omega-\Omega_o)/\sigma_{\Omega}$, where $\Omega_o/2\pi= 432$ nHz, at a
depth of $0.4\RSun$ and for different latitudes obtained from the inversion of
the different MDI $4\times 72$-day long data sets.}  Right panel: as in the
left panel, but for GONG data.  }
\label{fig:35}
\end{center}
\end{figure}

This result is intriguing, particularly if we analyze the time evolution of
the dip, for both MDI and GONG derived profiles, as shown in
Figure~\ref{fig:35}.  It was not possible to include the $1\times$ and
$2\times 72$-day long results, since the quality of the inverted profiles at
that depth and latitude is too low. Therefore, we used the $4\times 72$-day
long data, since its precision and temporal resolution allow us to carry out a
temporal evolution analysis with adequate quality of the resulting profiles.
{Although the dip is certainly at the limit of the resolution of the data
and the inversion method, there is a systematic temporal change of the dip.
This variation is not found at other latitudes.}

\begin{figure}[!ht]
\begin{center}
\includegraphics[width=.8218\columnwidth]{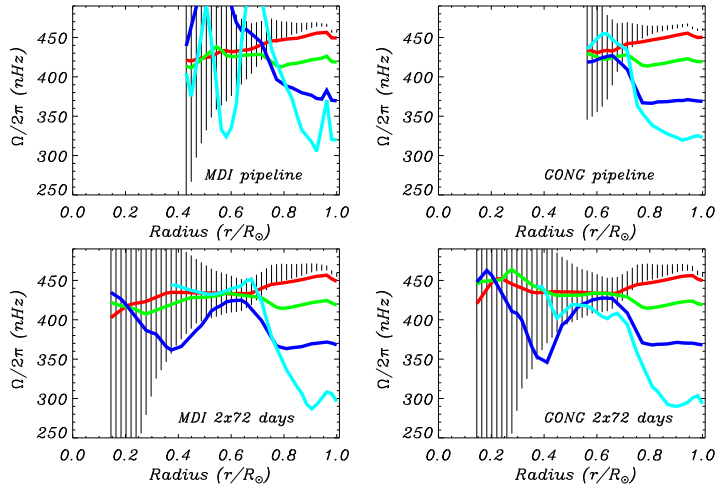}
\caption{Time-averaged rotational profiles obtained from the inversions of MDI
$2\times 72$, GONG $2\times 72$-day long, GONG pipeline, and MDI pipeline.
Black, red, green, dark-blue, and light-blue lines correspond to the rotational
rate at different latitudes, namely $0, 20, 40, 60$, and $80^{\rm o}$,
respectively. Vertical lines represent the error bars for the rotational rate
at the Equator.}
\label{fig:29}
\end{center}
\end{figure}

The consequences of using different peak fitting techniques on the inversion
results is illustrated in Figure~\ref{fig:29}. That figure shows the
time-averaged rotational rates obtained using MDI and GONG $2\times 72$-day
long alternative fitting method and GONG and MDI respective project
pipelines. The lengths of the fitted time-series are comparable, however the
spherical harmonic degree and frequency ranges of the fitted mode sets differ
significantly. In particular, the mode sets obtained by the project pipelines
result in rotational profiles that significantly disagree in the spatial
extent of the optimal inversion grid and in the inverted rotation rates at
high latitudes and in the radiative zone. The mode sets obtained through the
alternate technique devised for this work are, in contrast, homogeneous, even
though data from different instruments were fitted. Hence systematic
differences introduced by different fitting techniques and different mode sets
are greatly reduced.

\section{Conclusions}

We have fitted one solar cycle of MDI and GONG data and the latest HMI data
using a new fitting methodology. This method fits individual multiplets, an
asymmetric mode profile, incorporates all known instrumental distortion, uses
our best estimate of the leakage matrix, and uses an optimal sine
multi-tapered spectral estimator. It was applied to time series of varying
lengths to study the effect of trading off precision for temporal resolution in
the inversion results. On the other hand, the improved inversion method that we
used is one that estimates the optimal inversion model grid based on the
extent of the mode set (over spherical harmonic degree and frequency) and the
data uncertainties.

\begin{figure}[!ht]
\begin{center}
\includegraphics[width=.48\columnwidth]{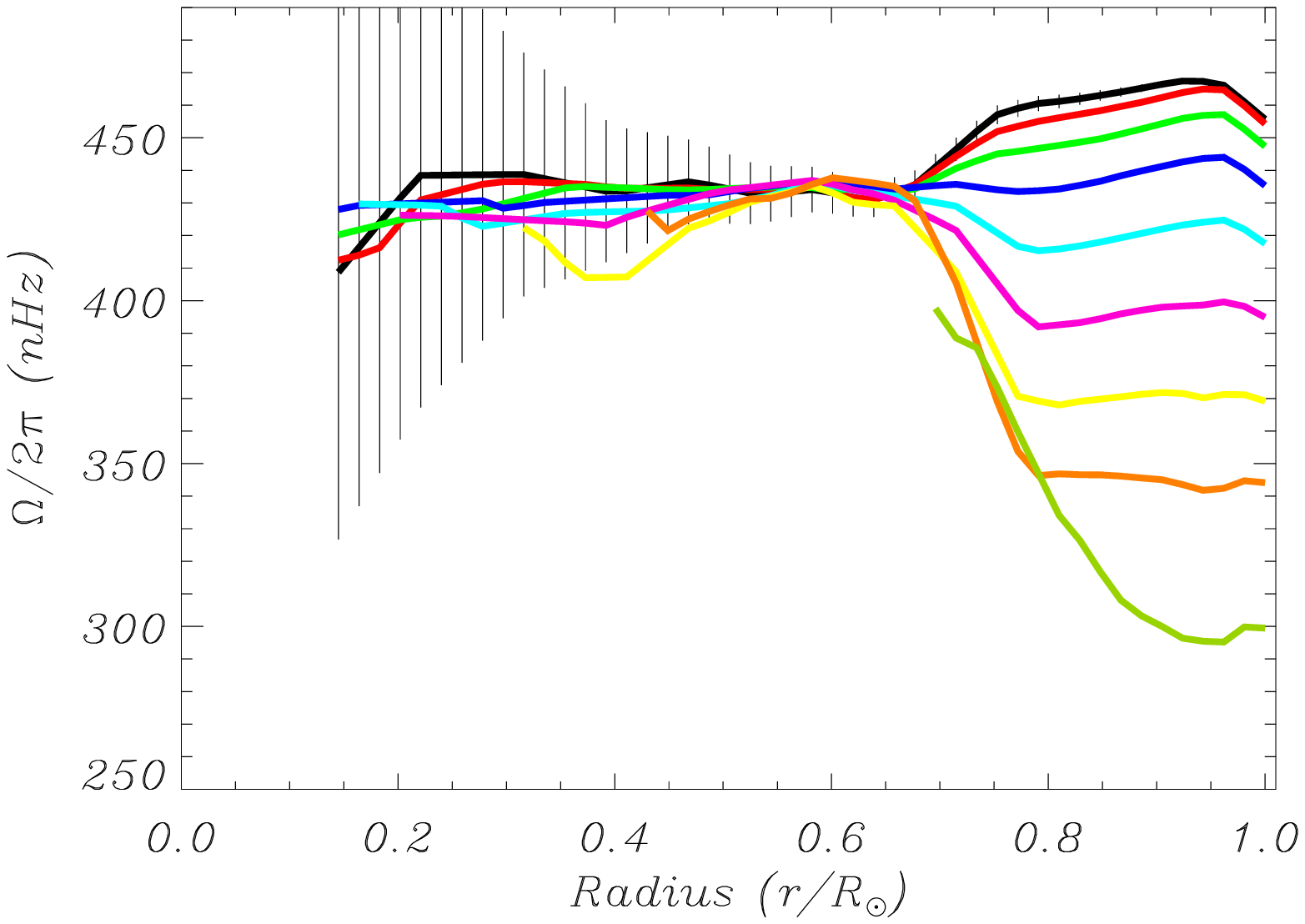}
\includegraphics[width=.48\columnwidth]{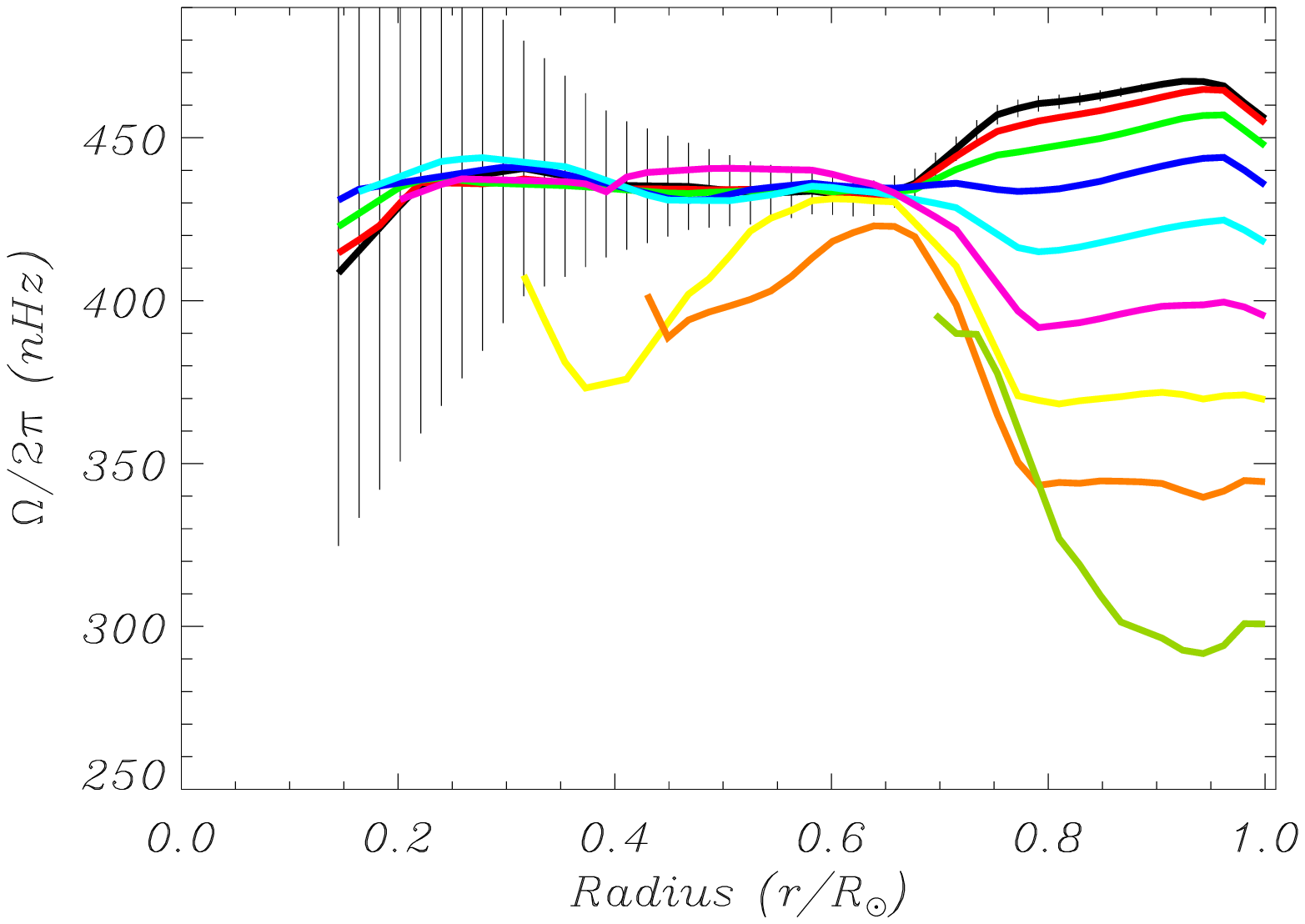}
\caption{Rotational profiles obtained from the inversions of MDI $64\times 72$
(left panel) and GONG $64\times 72$-day long (right panel).  The rotational
rates at different latitudes, from the Equator to $80^{\rm o}$ at steps of
$10^{\rm o}$, are represented by colored lines. Vertical lines represent the
error bars for the rotational rate at the Equator.}
\label{fig:18}
\end{center}
\end{figure}

Our results are summarized in Figure~\ref{fig:18}, where we present the
rotational profiles obtained from inverting frequency splitting derived from
fitting time series spanning an entire solar cycle, Cycle 23, for both GONG
and MDI observations.  These profiles are our best inferences of the rotation
in the radiative region, to date. Both results are compatible with a radiative
zone rotating rigidly at a rate of approximately 431 nHz; however, it is not
possible to disregard a faster or slower rotator below $0.2\RSun$ (\ie\  up to
600 or down to 300 nHz). Although the radiative zone seems to rotate rigidly,
there is a consistent and systematic dip in the rotation profile located at
around $0.4\RSun$ and $60^{\rm o}$ of latitude. This dip appears to evolve
with time, although this last result has to be confirmed when additional
time series covering Cycle 24 become available.

\end{article} 

\end{document}